# A Fabrication Method for Adaptive Dielectric Gradient Insulating Components

Zikui Shen[1,2], Zhidong Jia[2*], Yanpeng Hao[1], Zhenyu Xin[2], Xilin Wang[2]

[1] School of Electric Power Engineering, South China University of Technology, Guangzhou, 510630, China;
[2] Shenzhen International Graduate School, Tsinghua University, Shenzhen, 518055, China.
*jiazd@sz.tsinghua.edu.cn

**Abstract:** Dielectric gradient components have advantages in electric field mitigation and insulation improvement. In this paper, we propose a fabrication method for adaptive dielectric gradient components using in situ AC electric field, including the mechanism and the corresponding operational procedures for industrial applications. Based on the electric polarization and self-assembly effect of the filler particles in the liquid matrix, the chain-like structure in the high field strength region is constructed to enhance the local permittivity and mitigate the maximum of the spatial electric field. The dielectric gradient basin insulator is prepared by this method, and its flashover voltage is increased by 12.8% compared with that of a homogeneous dielectric basin insulator, and the improvement is 20.8% when metal particles are present on the surface. The more non-uniform the initial electric field is, the greater the improvement in flashover voltage. This method is expected to promote the industrial application of dielectric gradient insulating components.

## 1. Introduction

Non-uniform distribution of the electric field on insulation components increases the design difficulty and manufacturing cost of electrical equipment, and impedes the safe operation of the system[1–3]. Designing a uniform electric field on existing insulating components is beneficial to reduce the size of equipment, save environmental resources, and improve operational reliability, but it is also a recognized challenge[4, 5]. The conventional way of mitigating the electric field is mainly to optimise the topology[6], which has limited effect on electric field and increases the complexity of the structure. In recent years, researchers have been interested in dielectric gradient components, which combine material modification and topological optimization. The permittivity gradient in space can effectively mitigate the local concentration of electric field, improving the insulation performance with simple structure[7, 8].

Research on dielectric gradient insulating components began in the 1970's, and the current preparation methods include lamination[9], centrifugation[10], flexible casting[11], electrophoresis[12], magnetophoresis[13], 3D printing[14], and magnetron sputtering[15]. From the perspective of controlling, the lamination, flexible casting, 3D printing, and magnetron sputtering methods are constructing dielectric gradients by artificial "Building block", which cannot deal with the complex electric fields in space. While the centrifugal method, electrophoresis, magnetophoresis is to manipulate the filler particle distribution by applying an applied energy field to the liquid prepolymer insulating components. However, these energy fields are different from the electric fields in the physical laws. In terms of the particle distribution, only the magnetophoresis method could induce filler alignment, while all other methods are disordered. The disorderly distribution of fillers has a limited improvement of the effective permittivity of composites. In conclusion, the existing methods for preparing dielectric functional gradients have three problems: (1) the filler is randomly dispersed in the matrix, which limits the regulation range of the effective permittivity, and limits the application situations; (2) the resulting dielectric functional gradients cannot be perfectly matched with the actual electric field, either by artificial discrete control or by applied energy fields such as centrifugal force field, current field and magnetic field; (3) there are compatibility problems with the existing manufacturing process, resulting in high industrialization cost.

In view of this, a fabrication method of adaptive dielectric gradient insulating components is proposed in this paper. Self-contained electrodes are used to apply in-situ assisted AC electric field to the liquid prepolymer in the mold, constructing the adaptive dielectric gradient to mitigate the electric field subjected by insulating components during operation. We believe that this method can deal with the three problems above.

## 2. Theoretical background

### 2.1. Effective medium theory and electromechanics of particles

To find the relationship between fillers and the effective permittivity $\varepsilon_{eff}$ of composites, researchers have proposed several theoretical models based on tremendous experimental results, including the parallel model, series model, logarithmic model, and Maxwell–Garnett model. Among these models, the parallel model and series model are ideal and give the upper and lower limit of the $\varepsilon_{eff}$, respectively. The other models are based on random dispersion of fillers in the continuous phase matrix. Based on the permittivities of BaTiO$_3$ ($\varepsilon_p = 1235$) and epoxy resin ($\varepsilon_m = 4.5$), the values of $\varepsilon_{eff}/\varepsilon_m$ are calculated according to the abovementioned models. Experimental results from literature are inserted in Fig. 1[16–19]. The reported effective permittivities are rarely higher than the predicted values by the logarithmic model, the region between the parallel model and logarithmic model is missing[20].

As a result, increasing the $\varepsilon_{eff}$ of the composite by randomly dispersed fillers (even with extremely high permittivity) requires high filler load, which increases the viscosity of the compound and is detrimental to the molding process of insulating components, such as IGBT encapsulating materials[21]. Therefore, the parallel structure of filler and matrix must be constructed in the composite to significantly increase the $\varepsilon_{eff}$. In the field of functional materials, researchers have developed ice template method[22], electric field method[23, 24], and magnetic field method[25] to induce filler orientation to enhance specific properties such as elastic modulus and thermal conductivity, and electrical conductivity. This paper focuses on the effect of AC electric field on the liquid prepolymer composites.

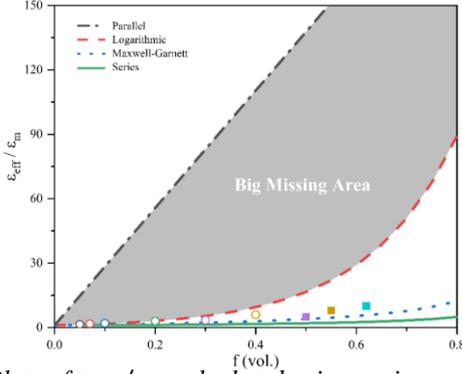

**Fig. 1.** *Plots of $\varepsilon_{eff}/\varepsilon_m$ calculated using various models for composites. Reproduced from*[20].

While dielectric parameters for particles and matrix are different, interfacial polarization would be generated in an AC electric field, called Maxwell-Wagner polarization. The polarized particles in the liquid matrix will be align in chains along the electric field, called the electric field self-assembly effect. This occurs frequently in electrical systems, for example, impurities in the insulating oil form "bridges" along the electric field. However, the assembly efficiency and the improvement of the effective permittivity of the composites after self-assembly differ for different shapes of filler particles. Appropriate filler shape helps to reduce the demand for assisted electric field and the required filling amount. Three common particle shapes are displayed in Fig. 2, including ellipsoids, spheres and rods.

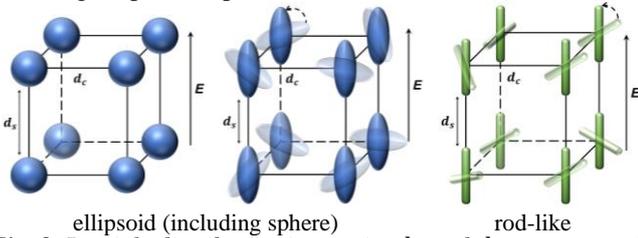

ellipsoid (including sphere)      rod-like

**Fig. 2.** *Particle distribution in matrix, $d_c$ and $d_s$ are core and surface distance of neighbouring particles, respectively.*

Assuming that the fillers are uniformly distributed in the matrix, the core distance of neighbouring filler is[26]

$$d_{cs} = 1.612R\varphi^{1/3} \tag{1}$$
$$d_{ce} = 1.612a(\lambda/\varphi)^{1/3} \tag{2}$$
$$d_{cr} = 0.923l(\lambda^2\varphi)^{-1/3} \tag{3}$$

where $d_{cs}$, $d_{ce}$ and $d_{cr}$ are core distances of ellipsoidal fillers and rod-like fillers, respectively, $R$ is the radius of the sphere, $a$ is semi-major axis of the ellipsoid, $l$ is the length of the rod, $\lambda$ is the aspect ratio of the particle (sphere is the ellipsoid with $\lambda$ of 1), $\varphi$ is the volume fraction of the fillers.

In the electric field, the suspended spherical dielectric particle is polarized and the dipole moment is[27]

$$p_s = 4\pi\varepsilon_m R^3 \left(\frac{\varepsilon_p-\varepsilon_m}{\varepsilon_p+2\varepsilon_m}\right)E_0 \tag{4}$$

For rod-like particles, the dipole moment can be calculated approximately based on the ellipsoidal particles (a>b=c). Decomposing the electric field $E_0$ in the parallel and perpendicular long axis, the induced dipole moments are[27]

$$\vec{p}_\parallel = \frac{4\pi ab^2}{3}\frac{\varepsilon_p-\varepsilon_m}{1+\left(\frac{\varepsilon_p-\varepsilon_m}{\varepsilon_m}\right)L_\parallel}E_0 cos\theta \tag{5}$$

$$\vec{p}_\perp = \frac{4\pi ab^2}{3}\frac{\varepsilon_p-\varepsilon_m}{1+\left(\frac{\varepsilon_p-\varepsilon_m}{\varepsilon_m}\right)L_\perp}E_0 sin\theta \tag{6}$$

where $\varepsilon_p$ and $\varepsilon_m$ are the permittivities of the particle and matrix, respectively, $\theta$ is the angle between major axis and electric field, $L_\parallel = \frac{b^2}{2a^2e^3}[\ln\left(\frac{1+e}{1-e}\right) - 2e]$, $e = \sqrt{1-b^2/a^2}$, $L_\perp = (1-L_\parallel)/2$.

If the dipole moment is not parallel to the electric field, an electric torque ($\vec{T} = \vec{p} \times \vec{E}$) is generated to drive the non-spherical particles to orient along the electric field direction. After orientation as shown in Fig. 2, the surface distance of neighbouring particles is[26]

$$d_{ss} = (1.612\varphi^{1/3} - 2)R \tag{7}$$
$$d_{se} = (1.612\lambda^{-\frac{2}{3}}\varphi^{-\frac{1}{3}} - 2)a \tag{8}$$
$$d_{sr} = (0.923\lambda^{-\frac{2}{3}}\varphi^{-\frac{1}{3}} - 1)l \tag{9}$$

When the volume fraction of the filler is high enough, the particles will contact each other after orientation. According to Eqs. (7-9), the threshold volume fraction is calculated as shown in Fig. 3, which is independent of the particle size and inversely related to the aspect ratio in a square way. As the aspect ratio increases, the threshold volume fraction decreases rapidly. For ellipsoidal particles, the threshold volume fractions are 52.4% and 0.524% when the aspect ratios are 1 (sphere) and 10, respectively. For rod particles, the values are 78.5% and 0.785%, when the aspect ratios are 1 (sphere) and 10, respectively.

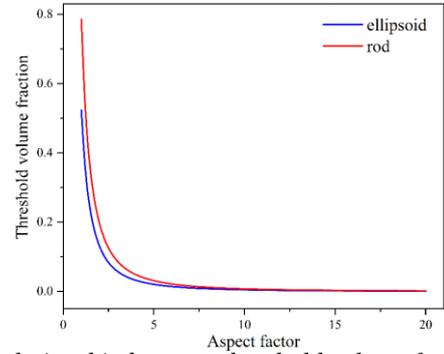

**Fig. 3.** *Relationship between threshold volume fraction ($d_s = 0$) and the aspect ratio of particles.*

The forces on the particle include the electrostatic force of neighbouring polarized particles and the drag force of the matrix. The attractive force between two neighbouring polarized spherical particles (colinear dipoles) is

$$F_{21} = -\frac{3p_s^2}{2\pi\varepsilon_m d_{cs}^4} \tag{10}$$

The attractive force decreases rapidly with core distance. The viscous drag subjected by spherical particles when translational moving in the liquid polymer is

$$F_{drag,s} = 6\pi r\eta v \tag{11}$$

where $\eta$ is the viscosity factor of the matrix, $v$ is the translational velocity of the particle. When the matrix viscosity, dielectric parameters of the filler, and applied electric field are identical, the smaller the surface distance, the higher the self-assembly efficiency. The detailed analysis for ellipsoidal and rod-like particles is shown in the supplementary materials Note 1.

The effect of filler shapes on effective permittivity improvement of the composites after fully self-assembly (as illustrated in Fig. 4) are evaluated. The improvement of ellipsoidal particles on the effective permittivity is independent of the aspect ratio and is lower than that of rod-like particles. The closer to the ideal cylinder, the higher the improvement effect of rod-like particles. The detailed derivation is shown in Note 2.

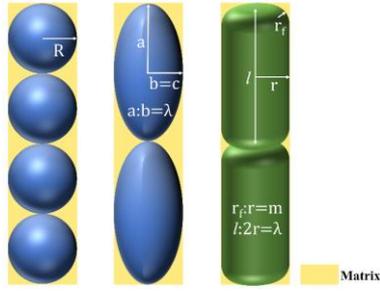

***Fig. 4.*** *Diagram of particles after fully assembly in the matrix.*

In summary, one-dimensional (1D) rod-like dielectric particles with high permittivity and high aspect ratio are preferred to be used as fillers in this method.

### 2.2. Transient and spatiotemporal analytical model of the effective permittivity of liquid composites

Fillers go from disorder to fully self-assembly in a time-domain process. The change of microstructure is inevitably mapped to macroscopic properties, and the previous work found that the effective permittivity of liquid composites will first increase and then stablize in a uniform AC electric field, typical results are shown in Fig. 5[28].

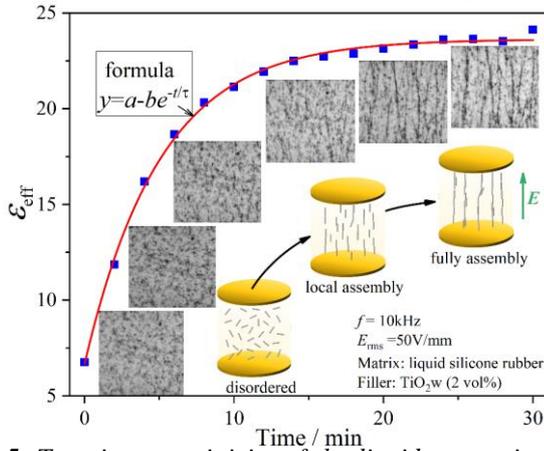

***Fig. 5.*** *Transient permittivity of the liquid composites in a uniform AC electric field, reproduced from*[28].

The transient process satisfies[28]
$$\varepsilon_{\text{eff}}(t) = a - be^{-\frac{t}{\tau}} \quad (12)$$
$$\tau = \alpha\varphi^{-1/2}E^{-2} \quad (13)$$
where $a-b$ and $a$ are the initial and stable permittivities, respectively, depending on material component, $\tau$ is the time constant, $\alpha$ is a constant, $\varphi$ is the filler volume fraction, and $E$ is the applied electric field strength, $t$ is the duration. The higher the applied field strength, the smaller the time constant and the faster the transient process of the permittivity.

The change in effective permittivity of the composite before and after curing satisfies[28]
$$\varepsilon_s = \varepsilon_l - \Delta\varepsilon \quad (14)$$
where $\Delta\varepsilon$ is due to the increase in cross-linkage, and depends only on the matrix materials.

The preparation steps for typical composite insulating components such as basin insulators, can be summarized as (1) pouring the liquid prepolymer composites into a casting mold and (2) waiting for material curing. If an AC voltage is applied through the self-contained electrodes in the first step, non-uniform electric field will be generated inside the liquid prepolymer composite component. According to the above theory, the speed of the transient process of permittivity varies in different areas[29]. According to Eq. 12 and Eq. 13, the kinetic equation for the evolution of the permittivity is[30]
$$d\varepsilon_{\text{eff}}/dt = \alpha^{-1}(a - \varepsilon_{\text{eff}})\varphi^{1/2}E^2 \quad (15)$$

The space electric field satisfies Maxwell's equations
$$\nabla \cdot D = 0 \quad (16)$$
$$D = \varepsilon_0\varepsilon_{\text{eff}}E \quad (17)$$
$$E = -\nabla\emptyset \quad (18)$$

Therefore, the permittivity and the electric field are coupled in time and space, and the above differential equations can be solved using finite element simulation software such as COMSOL. The time derivative of the permittivity at any point is proportional to the square of the instantaneous electric field strength. There exist such moments when the higher the local field strength, the higher the local permittivity, which happens to be the target of the dielectric gradient components.

As mentioned earlier, we propose for the first time that constructing dielectric gradient by the in-situ AC electric field induction (similar to the actual operating electric field) in the liquid prepolymer stage during the preparation of the insulating components. This method is inspired from "insects versus insects" in "*Brush Talks from Dream Brook*", and follows the wisdom of "give someone a taste of his own medicine". The detailed mechanism is described in Fig. 6.

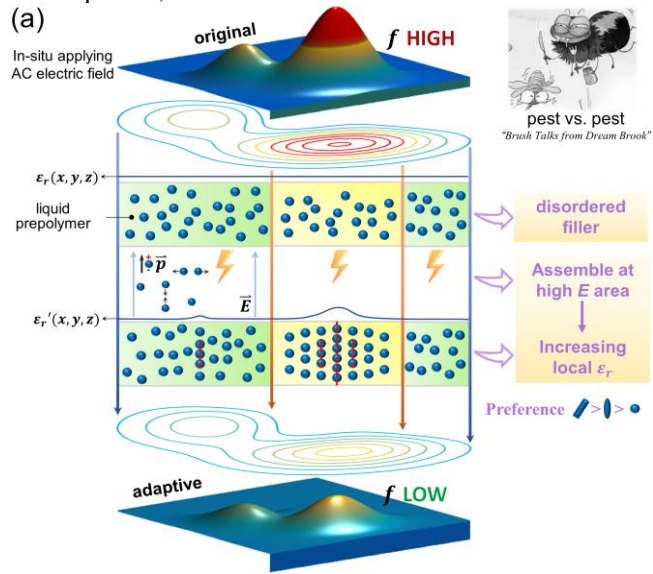

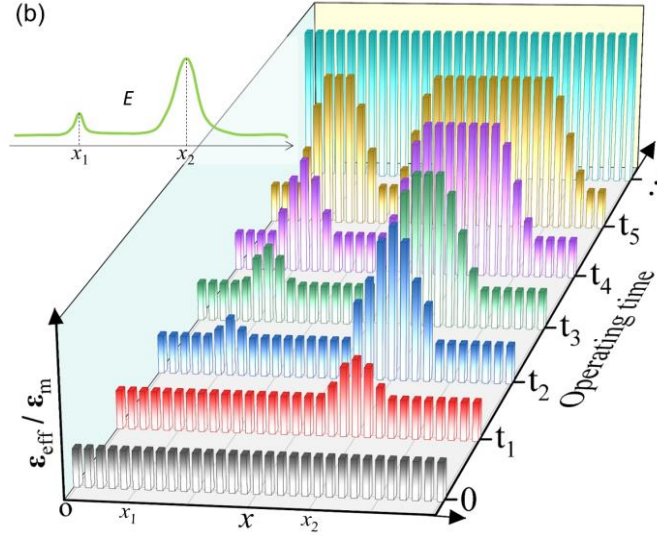

*Fig. 6. The mechanism of this method, inspired by "insect vs. insect" in "Brush Talks from Dream Brook".*
*(a) 3D schematic of the permittivity-electric field evolution, (b) 1D illustration of the permittivity evolution over time.*

Fig. 6a shows a 3D schematic of the permittivity-electric field evolution. Without artificial search, the filler particles in the high electric field region are always the first to assemble in chains, increasing the local permittivity to mitigate the maximum of the spatial electric field. Fig. 6b shows a 1D illustration of the permittivity evolution over time. Assuming that there are two peaks of the 1D electric field in the *x*-axis (green curve in the figure), the permittivity of the liquid prepolymer is $\varepsilon_i$ (close to $\varepsilon_m$) at the initial time ($t = 0$); at time $t_1$, the permittivity near $x_2$ is the first to increase, and the other area do not change; at time $t_2$, the permittivity near $x_1$ starts to increase; at time $t_3$, the permittivity near $x_2$ reaches the stable value; at time $t_4$, the permittivity near $x_1$ also reaches the stable value. If the electric field is applied for a long enough time, the permittivity increases to the stable value in all regions. In this time series, not all the permittivity distribution is favourable to mitigate the electric field, which has been verified in Ref. 30 based on finite element simulation. It is necessary to remove the assisted electric field at the right time (such as $t_3$) and then "freeze" (curing) the resulting dielectric gradient.

## 3. Preparation process design

The simulation analysis in the Ref. 30 defaults to three assumptions: (1) the viscosity of the matrix is constant during the application of the assisted voltage; (2) ignoring gravity, the filler particles do not settle; and (3) ignoring thermal motion of the particles, the constructed dielectric gradient remains stable during the curing process. If the above three assumptions are valid, the conditions that need to be met are

(1) Ensure that the viscosity of the matrix changes smoothly. Modify the crosslinking speed of the matrix material by adjusting the amount of the accelerator (or inhibitor). Control assistant time at 3~10 min (author recommends). Increasing the amplitude of the power supply will reduce the assistant time. If the assistant time is too short, it will amplify the human operational error;

(2) Check the viscosity of the matrix at 2~100 Pa·s (recommended by authors). If the viscosity is too low, add reinforcing filler (such as silica, etc.) or pre-cure to increase the viscosity to prevent the target filler from settling; conversely, increase the temperature to reduce the viscosity to avoid the assisted voltage amplitude required is too high.

(3) Continue to apply the balance voltage, after removing the assisted voltage, to prevent the self-assembly chain from being disrupted until the viscosity of the matrix exceeds 100 Pa·s or gelatin. Destabilisation factors include both the thermal motion of the particles and the convection of the low viscosity matrix due to uneven heating. Microscopic particles follow the Boltzmann distribution in the free state. When the applied electric field is greater than a certain threshold, the particle surface spacing region tends to 0 by solving the Boltzmann distribution equation, at which time the applied electric field can suppress the damage on the self-assembly chain by thermal motion. It has been found that the potential energy difference between the particles before and after chain formation must be greater than $9kT$ for stable "pearl chains" to be observed[31]. For spheric particles, the expression of the threshold electric field strength is

$$E_{\text{th}} = \frac{1.7}{R^{1.5}|\text{CM}|}\sqrt{\frac{kT}{\varepsilon_0\varepsilon_m}} \quad (19)$$

where $k$ is the Boltzmann constant, $\text{CM} = (\varepsilon_p - \varepsilon_m)/(\varepsilon_p + 2\varepsilon_m)$ is the Clausius-Mossotti factor. The smaller the equivalent particle size of the filler, the smaller the CM factor or the higher the temperature, the greater the field strength required to maintain the chain. Using $R$=1 μm, CM=1, $\varepsilon_m$=15, and $T$=300 K as an example, the threshold field strength is calculated to be 9 V/mm. For rod-like or ellipsoidal particles, the threshold field strength is[32]

$$E_{\text{th}} = \frac{7.7}{l^{1.5}|\text{CM}|}\sqrt{\frac{kT}{\varepsilon_0\varepsilon_m}} \quad (20)$$

Using $l$=5 μm, CM=1, $\varepsilon_m$=15, and $T$=300 K as an example, the threshold field strength is 0.4 V/mm. When choosing the balancing voltage, it is necessary to ensure that the field strength generated in the chain assembly region exceeds the threshold value. At the same time, it is much smaller than the assisted voltage so as not to affect the already existing dielectric gradient (e.g., to continue to induce filler assembly), and the author recommends that the difference between the assisted voltage and the balance voltage be more than 10 times (corresponding to a difference of more than 100 times in the transient process time constant).

In summary, the overall workflow of this method is shown in Fig. 7.

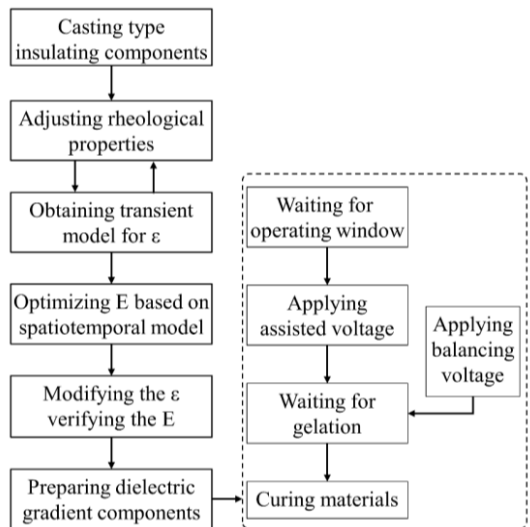

*Fig. 7. Overall workflow of the method.*

## 4. Results and Discussions

The dielectric functional gradient basin insulator was prepared based on this method, and microscopic observation and insulation performance tests were performed. Fig. 8 shows the casting setup of the scaled basin insulator. The basin insulator is 20 mm high and 120 mm outside diameter, with an initial electric field non-uniform factor of 3.5. The differences from the existing molding process are: (1) in addition to the mold and the insert electrodes, a conductive rod and a cylindrical ground electrode are added to generate an assisted electric field similar to the actual operation electric field; (2) the material of the mold is replaced from metal to PTFE to eliminate electromagnetic shielding to the molding cavity.

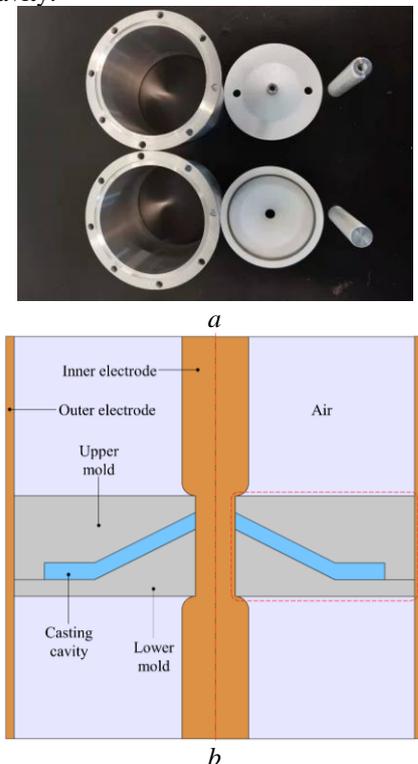

*Fig. 8. Casting setup of the basin insulator.*
*(a) casting mold and (b) section schematic.*

### 4.1. Rheological properties optimization

The $TiO_2w/Al_2O_3/E39$ epoxy resin system was selected, where the content of $Al_2O_3$ was 100 phr. The effect of the accelerator dosage and pre-curing time on the matrix viscosity was investigated. E39 and MeHHPA were mixed in the ratio of 100:66 by mass, and then 1, 2, 3 and 4 phr of BDMA accelerator (relative to E39) were added and stirred well, next the viscosity of the mixture was measured using a rotational viscometer NJ-1. Fig. 9 shows the variation of the mixture viscosity at 25 °C with time.

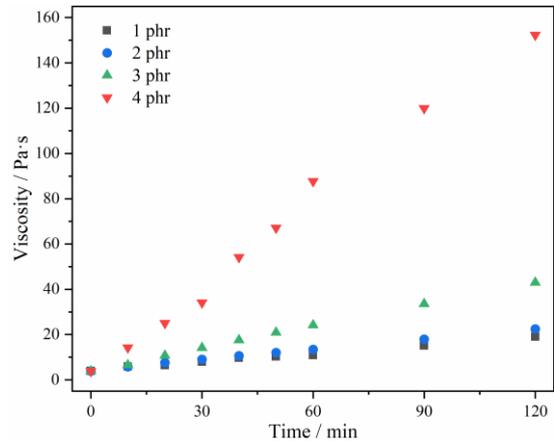

*Fig. 9. Variation of the of the mixture viscosity at 25 °C.*

The initial viscosity of the mixture was about 3.9 Pa·s, and the viscosity increased with time. While 1 or 2 phr accelerators were added, the viscosity changed slowly, and the viscosity after 2 h was 19 and 22 Pa·s, respectively; while 3 phr or 4 phr accelerators were added, the viscosity after 2 h was 43 Pa·s and 152 Pa·s.

When the accelerator content is more than 3 phr, the viscosity of the matrix changes quickly, which is not suitable for accurate control; when the accelerator content is 1 phr, it takes a long time to increase the viscosity to prevent the filler from settling. Therefore, accelerator content of 2 phr was selected. The initial electric field non-uniformity coefficient of this structure is 3.5, and the preferred compatibility coefficient is 1 according to Ref. 30. Therefore, 2 vol% (approximately 15 phr) of $TiO_2w$ should be added, and the basin insulator casting compound is shown in Table 1.

**Table 1.** Casting compound (phr) for basin insulator

|  | E39 | MeHHPA | BDMA | $Al_2O_3$ | $TiO_2w$ |
|---|---|---|---|---|---|
| **Content** | 100 | 66 | 2 | 100 | 15 |

The transient permittivity of the mixture was evaluated to determine the window of operation. Mixtures with 2 phr accelerator were pre-cured for 0, 1, 2 and 3 h. Using the device in Ref. 28, the liquid mixture was poured into a PTFE ring spacer (inner diameter of 30 mm and thickness of 2 mm) between two parallel plate electrodes. Transient permittivity was measured by applying an electric field of 10 kHz/50 V/mm to mixtures with different levels of precuration. The transient permittivity of the mixture is shown in Fig. 10, and the fitted parameters of the transient process are extracted in Table 2.

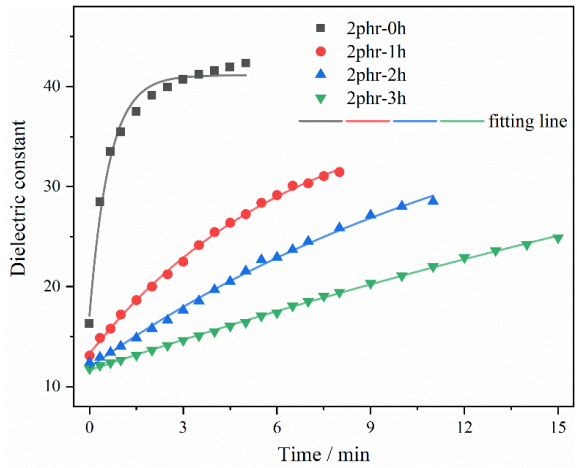

*Fig. 10. Transient effective permittivity of the mixture.*

**Table 2.** Fitted parameters of the transient permittivity

| Samples | Electric field (V/mm) | Time constant (s) | Initial $\varepsilon_0$ | $\Delta\varepsilon_0$ |
|---|---|---|---|---|
| 2phr-0h | 50 | 37 | 16.3 | / |
| 2phr-1h | 50 | 393 | 13.1 | -3.2 |
| 2phr-2h | 50 | 864 | 12.4 | -3.9 |
| 2phr-3h | 50 | 2806 | 11.8 | -4.5 |

As the pre-curing time increases, the initial value and change rate of the permittivity both decrease significantly. The former is due to the increased curing degree of the matrix, and the latter is due to the increased viscosity of the matrix. After adding 2 phr accelerator (0~3 h), the time constant of the permittivity transient process increased from 37 s to 2806 s, and the change in the initial value of permittivity was 4.5. Since the initial value of permittivity changed from the stable value, Δε should be corrected in Eq. 14.

From the point of view of operational feasibility, assisted voltage should be controlled in a suitable range (e.g. < 10 kV) and the power supply of the preparation process should not be overly demanding. In summary, 2 phr accelerator was added and pre-cured for 2 h before the electric field was applied to construct the dielectric gradient.

### 4.2. Electric field optimization based on spatiotemporal model

The above experimental data were substituted into the spatiotemporal model for simulation to optimize the electric field distribution. The optimized result is an assisted voltage RMS of 1.5 kV and a duration of 200 s, which can reduce the maximum field strength of the concave triple junction to 80%.

All the above analyses are based on liquid prepolymer composites in the casting process, and there are still differences with the real operating basin insulators: (1) there is no PTFE mold in the real operation and the surrounding is filled with insulating gas; (2) there are differences in the permittivities of the composites before and after curing. Therefore, secondary processing of the simulation results is necessary.

First, the permittivity of the mold in the simulation program is set to 1, i.e., air or $SF_6$ insulating gas. Second, correct the permittivity of the basin insulator according to Eq. 14. In COMSOL, the distribution of permittivities of liquid prepolymer composites corresponding to the optimal state of the electric field is exported as a text file (.txt), which contains the coordinate information and corresponding permittivities; in the new project file, the above text file data are imported and the new function is defined by interpolation, with the independent variable as spatial coordinates and the dependent variable as permittivities. According to Eq.14 and using the interpolation function to define the permittivity in the domain of the solid basin insulator (for the composites in this work, Δε is 8). Fig. 11 shows the results of the static electric field simulation.

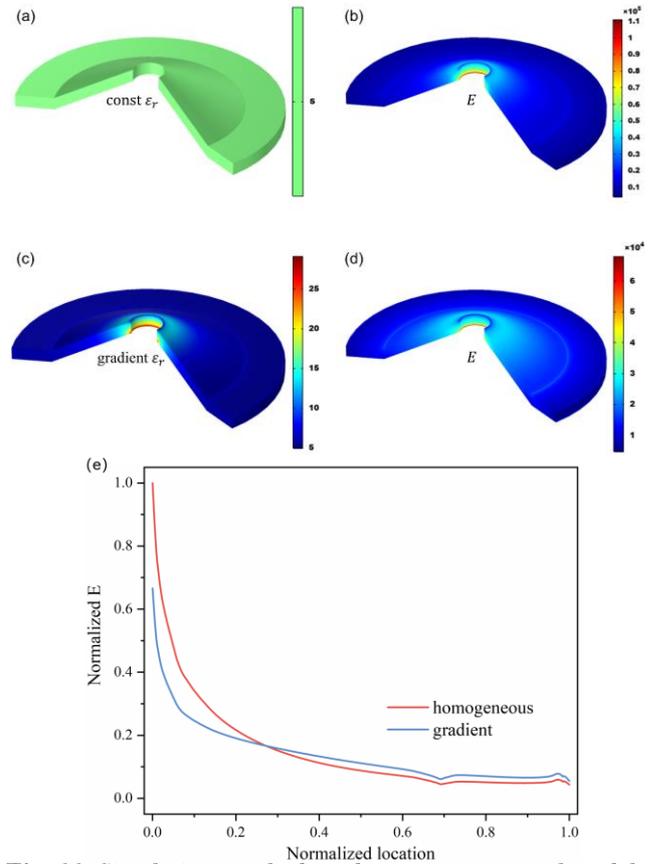

*Fig. 11. Simulation results based on spatiotemporal model. (a-b) constant permittivity and corresponding electric field distribution of the basin insulator (the voltage is 1 kV), (c-d) optimal gradient permittivity and the corresponding electric field distribution, (e) electric field strength distribution along the concave surface.*

Figs. 11a-b show the constant permittivity and corresponding electric field in the basin insulator (the voltage is 1 kV). Figs. 10c-d show the optimal gradient permittivity constructed by above method and the corresponding electric field. Fig. 11e compares the distribution of electric field on the gas side of the concave surface of the basin insulator for both cases. Compared to the homogeneous material, the electric field strength at the triple junction on the concave surface of the permittivity gradient insulator is reduced by 32%. In the left 25% region, the surface electric field strength is weakened; in the right 75% region, the surface electric field strength is slightly higher than that of the homogeneous dielectric basin insulator. The dielectric gradient achieves the transfer of electric field stress apportionment from the inner electrode to the ground electrode.

Fig. 12 shows the apparatus for electric field-assisted preparation of dielectric functional gradient basin insulators, which includes a power supply module, a control module, a casting mold and a vacuum oven.

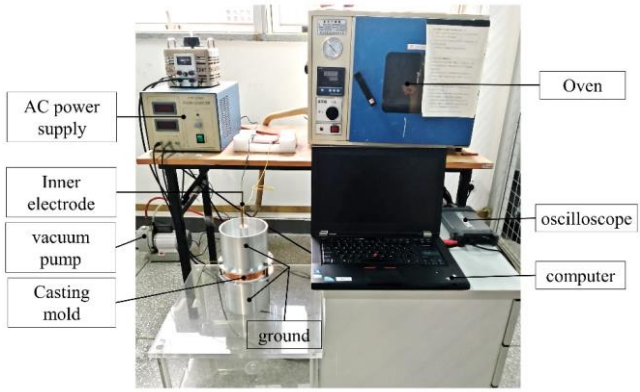

***Fig. 12.*** *Apparatus for electric field-assisted preparation of dielectric gradient basin insulators.*

Based on this apparatus, 2 sets of basin insulators were prepared separately. Sample 1 was prepared without electric field assistance, while sample 2 with electric field assistance. The insulator preparation process is shown in Fig. 13.

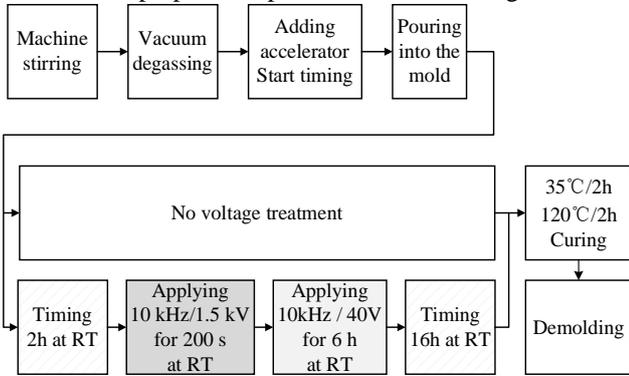

***Fig. 13.*** *Preparation processes of basin insulator. RT is room temperature.*

### 4.3. Characterization and Testing

The basin insulator was sliced at the three locations marked in Fig. 14a, and the dispersion state of the filler was observed using electron microscopy.

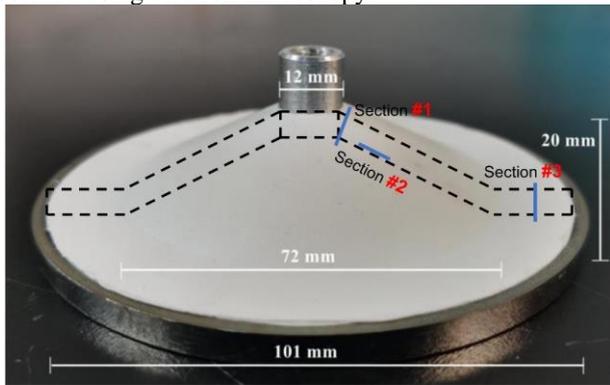

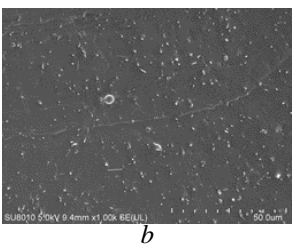

*b*

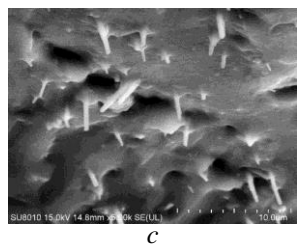

*c*

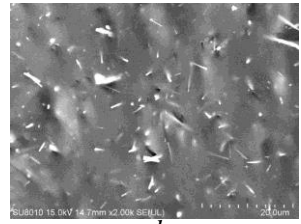

*d*

***Fig. 14.*** *Results of microscopic observation.*
***(a)*** *basin insulator and section location,*
***(b-d)*** *filler distribution in sections #1, #2, and #3.*

The simulation results show that the initial electric field strength at section #1 and section #2 is higher and the improvement of permittivity is larger, and the electric field line at section #1 is perpendicular to the profile and at section #2 is approximately parallel to the profile; the electric field strength at section #3 is weaker and the permittivity is almost unchanged. The $TiO_2$ws perpendicular to the paper are observed at section #1; the $TiO_2$ws parallel to the paper are observed at section #2; the randomly distributed $TiO_2$ws are observed at section #3. These are consistent with the simulation results.

The flashover voltage of the sample was tested in 4 states: state 1 is clean surface and other states are with metal particles in different locations. Fig. 15a illustrates the location of metal particles on the concave surface of the basin insulator, where the length of the metal particles is 2 mm.

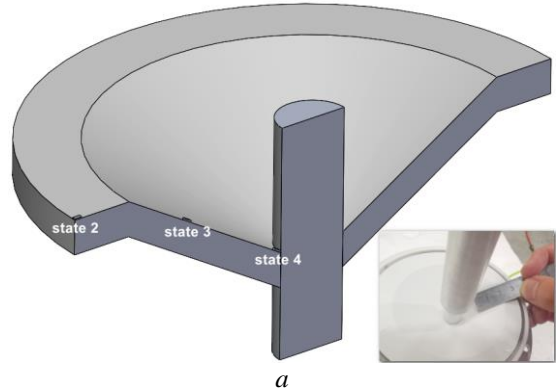

*a*

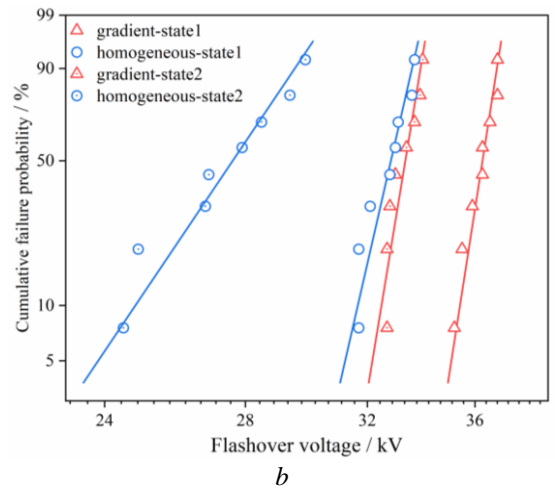

*b*

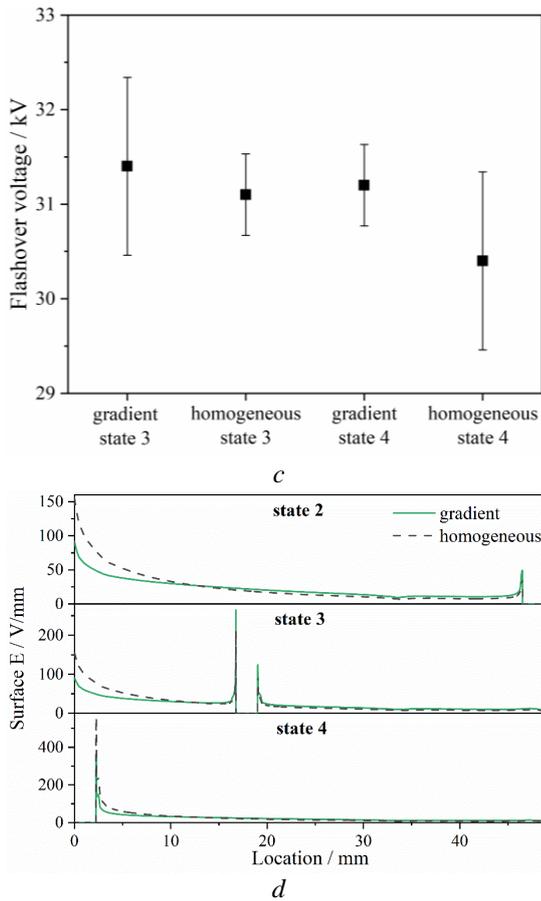

*Fig. 15. Results of insulation test.*
*(a) illustration of the metal particles on the concave surface,*
*(b-c) results of the flashover voltage,*
*(d) electric field distribution along the concave surface.*

In state 1, the probable flashover voltage is 32.6 kV for sample 1 and 36.6 kV for sample 2, which is 12.4% higher than the former (simulations show that the maximum field strength at the triple bond point decreases by about 25%); in state 2, the probable flashover voltage drops to 27.8 kV for sample 1 and 33.6 kV for sample 2, which is 20.8% higher than the former, as shown in Fig. 15b.

In state 3 and state 4, the flashover voltages of both are similar, as shown in Fig. 15c. Fig. 15d shows the distribution of the electric field strength with the radius on the concave side when the metal particles are in different locations. In state 2, due to the shielding effect of the cylindrical ground electrode, the maximum electric field strength along the surface at this time is still located at the triple junction point of the inner electrode. In state 3, the distortion effect on the field strength is obvious, and the maximum electric field strength of the concave surface of both groups of insulators is shifted to the side of the metal particles near the inner electrode, and is much higher than the triple junction point of the inner electrode. The effect of the distorted electric field is most severe when the metal particles are located near the inner electrode. Compared with the uniform medium, the electric field distribution of dielectric gradient basin insulators is improved in all three cases, however, the electric field distortion is severe in the latter two cases so that the flashover voltages are similar.

## 5. Conclusion

This paper focuses on a fabrication method for adaptive dielectric gradient insulating components. Theoretical analysis, simulation and experiments were carried out, leading to the following conclusions.

(1) Compare to the applied electric field strength, the dielectric difference between the particles and matrix, the matrix viscosity and the filler content, increasing the aspect ratio of the filler is the most effective way to improve the efficiency of electric field-induced self-assembly.

(2) The parallel structure with the matrix after self-assembly will significantly increase the effective permittivity of the composite. The improvement of ellipsoidal particles on the effective permittivity is independent of the aspect ratio and is lower than that of rod-like particles. The closer to the ideal cylinder, the higher the improvement effect of rod-like particles. Considering the self-assembly efficiency and the effective permittivity, the 1D rod-like dielectric fillers with high permittivity and high aspect ratio are preferred to use in this method.

(3) The reference operation procedure of the method for industrial applications is proposed, and suggested values are given including matrix viscosity, parameters of fillers, filler content, operation window, operation time, and assisted voltage amplitude. Importantly, the balancing voltage is introduced to counteract the disruption of the dielectric gradient by thermodynamic motions.

(4) Based on this method, the scaled-down basin insulator was prepared. Rheological properties of the $Al_2O_3/TiO_2w/E39$ system were studied, and the suitable operating window was found, which provides the guidelines for the epoxy resin system. Microscopic observations show that the $TiO_2w$ alignment site agrees with the simulation results and the flashover voltage of the dielectric gradient insulator is increased by 12.4%. When metal particles are present near the ground electrode, the flashover voltage of the dielectric gradient insulator is enhanced by 20.8%.

In summary, this method is expected to promote the industrial application of dielectric functional gradient insulating components, broaden the application scope of high voltage technology, and transform the preparation process of insulating materials. In the future, to measure the permittivity distribution of dielectric gradient insulators non-destructively may be a worthwhile research issue.

## 6. Acknowledgments

This work was supported by the National Natural Science Foundation of China (No. 52177021) and Shenzhen fundamental research and discipline layout project (No. JCYJ20180508152044145).